\documentclass[aps,pra,preprint,amsmath]{revtex4-1}
\usepackage{silence}
\usepackage{mathrsfs}
\usepackage{graphicx}
\usepackage{epstopdf}
\usepackage{amsthm}
\usepackage{amsmath}
\usepackage{amsfonts}
\usepackage{amssymb}
\usepackage{braket}
\usepackage{bbold}
\usepackage{subfigure}
\usepackage{appendix}
\usepackage{tikz}
\usepackage{bm}
\usepackage{float}
\def\be{\begin{equation}}
\def\ee{\end{equation}}
\def\ba{\begin{array}}
\def\ea{\end{array}}

\def\1{{\bf{1}}}

\input amssym.def
\begin{document}
\title{\bf Summation and product forms of uncertainty relations based on metric-adjusted skew information}

\author{Cong Xu$^1$}
\author{Qing-Hua Zhang$^2$}
\author{Shao-Ming Fei$^{1,3}$}
\email{feishm@cnu.edu.cn}
\affiliation{
$^1$School of Mathematical Sciences, Capital Normal University, Beijing 100048, China\\
$^2$School of Mathematics and Statistics, Changsha University of Science and Technology, Changsha 410114, China\\
$^3$Max-Planck-Institute for Mathematics in the Sciences, 04103 Leipzig, Germany
}

\begin{abstract}
Uncertainty principle is one of the most essential features in quantum mechanics and plays profound roles in quantum information processing. We establish tighter summation form uncertainty relations based on metric-adjusted skew information via operator representation of observables, which improve the existing results. By using the methodologies of sampling coordinates of observables, we also present tighter product form uncertainty relations. Detailed examples are given to illustrate the advantages of our uncertainty relations.

\end{abstract}

\maketitle

\noindent {\bf 1. Introduction}

Since the uncertainty relation proposed by Heisenberg \cite{HW} in $1927$, Robertson \cite{RH} derived the well-known product form uncertainty relation satisfied by two arbitrary observables $A$ and $B$ with respect to a quantum state $|\psi\rangle$,
\begin{equation}\label{eq1}
\Delta A\Delta B\geq \frac{1}{2}|\langle\psi|[A,B]|\psi\rangle|,
\end{equation}
where $[A,B]=AB-BA$ and $\Delta M=\sqrt{\langle\psi|{M}^2|\psi\rangle-{\langle\psi|M|\psi\rangle}^2}$
is the standard deviation of an observable $M$. Since then many uncertainty relations have presented in terms of different approaches and quantities such as entropy
\cite{DD,MHU,WSWA,RAE}, variance \cite{GUDDER,DD1,DD2,SL}, successive measurements \cite{SMD}, majorization techniques \cite{PZRL,RLPZ,RL,FSGV} and skew information \cite{LUO5}.

In $1963$, Wigner and Yanase \cite{WY} introduced the skew information $I_{\rho}(A)$ to quantify the non-commutativeness of an observable $A$ associated with a  quantum state $\rho$,
\begin{align}\label{eq2}
I_{\rho}(A)=-\frac{1}{2}\mathrm{Tr}
\left(\left[\sqrt{\rho},A\right]^2\right)
=\frac{1}{2}\left\|\left[\sqrt{\rho},A\right]\right\|^{2},
\end{align}
where $\|\cdot\|$ denotes the Hilbert Schmidt norm. For pure states, the Wigner-Yanase (WY) skew information coincides with the variance \cite{LUO3}. A more general quantity introduced by Dyson is the Wigner-Yanase-Dyson (WYD) skew information,
\begin{align}\label{eq3}
I_{\rho}^{\alpha}(A)
=&-\frac{1}{2}\mathrm{Tr}[\rho^\alpha,A][\rho^{1-\alpha},A]\notag \\
=&\mathrm{Tr}(\rho A^2)-\mathrm{Tr}(A\rho^{\alpha}A\rho^{1-\alpha}) ,\,\,0\leq \alpha \leq 1.
\end{align}
Lieb \cite{Lie1} proved that the WYD skew information is convex with respect to $\rho$, a fact used to verify the monotonicity of quantum relative entropy and the strong subadditivity of the quantum entropy \cite{Lie2, Lin1,Lin2}.

The quantum Fisher information is initially defined by \cite{Hel},
\begin{align}\label{eq4}
I_{\rho}^{F}(A)=\frac{1}{4}\mathrm{Tr}(\rho L^2),
\end{align}
where $L$ is the solution of the operator equation $(L\rho+\rho L)=2\mathrm{i}[\rho,A]$.
In $1996$, Petz \cite{PED} introduced the concept of monotone metric on matrix spaces. By using operator monotone metrics, Hansen \cite{HF} defined a class of generalized quantum Fisher information called metric-adjusted skew information, which includes the WY skew information and the WYD skew information as special cases.

The summation form uncertainty relations play essential role in detecting quantum entanglement \cite{HHF,LL}. To this end, the summation form of quantum Fisher information defined by the symmetric logarithmic derivative could be superior to that of WY information \cite{LL}, as given by the quantum Cram$\acute{e}$r-Rao inequality. The summation form uncertainty relations via WY skew information and generalized skew information for observables and quantum channels have attracted considerable attention and a host of results have been obtained \cite{YANA1,YANA2,WU2,XWF1,XWF2,XWF3,CB3,FSS,ZL,ZWF1,ZWF2,CAL,RRNL,HLTG,HWF1}. In \cite{CB3}, Chen et al. proposed summation uncertainty for WY skew information. Cai \cite{CAL} confirmed that the results in \cite{CB3} also hold for all metric-adjusted skew information. Recently, product and summation forms of uncertainty relations have been introduced in \cite{MZF,HJ} for metric-adjusted skew information by using operator representation of observables and refinements of the Cauchy-Schwarz inequality.

The remainder of this paper is structured as follows. In Section 2, we first recall the definition of the metric-adjusted skew information. We establish the tighter summations form uncertainty relations of observables with respect to metric-adjusted skew information. The product form uncertainty relation via metric-adjusted skew information are presented in Section 3 and we conclude with a summary in Section 4.

\noindent {\bf 2. Sum uncertainty relations via metric-adjusted skew information}

Let $M_d(\mathbb{C})$ and $\mathcal{U}_d$  be the set of all complex $d\times d$ matrices and positive semi-definite matrices with trace $1$, respectively. For arbitrary $A,B\in M_d(\mathbb{C})$, $\rho \in \mathcal{U}_n$, $K_{\rho}(\cdot,\cdot):M_n(\mathbb{C})\times M_d(\mathbb{C})\longmapsto\mathbb{C}$
is called as symmetric monotone metric \cite{PED}, which can be written as
\begin{align}\label{eq5}
K_{\rho}(A,B)=\mathrm{Tr}[A^{\dag}c(L_{\rho},R_{\rho})B],
\end{align}
where $L_{\rho}(A)=\rho A$ and $R_{\rho}(A)=A\rho$ are the left and right multiplication operators such that $[\rho,A]=\rho A-A\rho=(L_{\rho}-R_{\rho})A$. $c(L_{\rho},R_{\rho})$ is called Morozova-Chentsov function taking the form
\begin{align}\label{eq6}
c(x,y)=\frac{1}{yf(xy^{-1})}, \,\,\,\, x,y>0,
\end{align}
where $f:R_+\longmapsto R_+$ is operator monotone ($R_+$ is the set of all positive real number), namely, if $A\geq B$ then $f(A)\geq f(B)$ for any $A,B>0$, and $f(t)=tf(t^{-1})$ for arbitrary $t>0$. Petz \cite{PED} proved that there exists a one to one correspondence between the symmetric monotone metrics $K_{\rho}(\cdot,\cdot)$ and the set of functions $f$.

In \cite{HF} Hansen introduced the metric-adjusted skew information,
\begin{align}\label{eq7}
I_{\rho}^{c}(A)
=&\frac{m(c)}{2}K_{\rho}^{c}(\mathrm{i}[\rho,A],\mathrm{i}[\rho,A])\notag \\
=&\frac{m(c)}{2}\mathrm{Tr}\{\mathrm{i}[\rho,A]c(L_\rho,R_\rho)\mathrm{i}[\rho,A]\},
\end{align}
where $c(x,y)$ is a regular metric \cite{HF}. Since $\mathrm{i}[\rho,A]=\mathrm{i}(L_\rho-R_\rho)A$, (\ref{eq9}) can be written as
\begin{align}\label{eq8}
I_{\rho}^{c}(A)=\frac{m(c)}{2}\mathrm{Tr}\{A\widehat{c}(L_\rho,R_\rho)A\},
\end{align}
where $\widehat{c}(x,y)=(x-y)^2c(x,y)$, $x,y>0$.

The metric-adjusted correlation of observables $A$ and $B$ is given by \cite{MZF,HJ}
\begin{align}\label{eq9}
Corr_{\rho}^{c}(A,B)
=&\frac{m(c)}{2}K_{\rho}^{c}(\mathrm{i}[\rho,A],\mathrm{i}[\rho,B]).
\end{align}
If one considers
\begin{align}\label{eq10}
f_{\alpha}(t)=\alpha(1-\alpha)\frac{(1-t)^2}{(1-t^\alpha)(1-t^{1-\alpha})},\,\,\,   t>0,
\end{align}
with $f_{\alpha}(0)=\alpha(1-\alpha)$, the corresponding Morozova-Chentsov function is
\begin{align}\label{eq11}
c_{\alpha}(x,y)=\frac{1}{\alpha(1-\alpha)}\frac{(x^\alpha-y^\alpha)
(x^{1-\alpha}-y^{1-\alpha})}{(x-y)^2},~ 0<\alpha<1.
\end{align}
And the associated monotone metrics is
$K_{\rho}^{\alpha}(A,B)=\mathrm{Tr}\{A^{\dag}c_{\alpha}(L_\rho,R_\rho)B\}$.
When $c(x,y)=c_{\alpha}(x,y)$, the metric-adjusted skew information reduces to the WYD skew information,
\begin{align}\label{eq12}
I_{\rho}^{c_{\alpha}}(A)
=&\frac{\alpha(1-\alpha)}{2}\mathrm{Tr}(\mathrm{i}[\rho,A]
c_{\alpha}(L_\rho,R_\rho)\mathrm{i}[\rho,A])\notag \\
=&-\frac{1}{2}\mathrm{Tr}\left[\rho^\alpha,A\right]
\left[\rho^{1-\alpha},A\right]=I_{\rho}^{\alpha}(A).
\end{align}
In particular, $I_{\rho}^{c_{\alpha}}(A)$ reduces to the WY skew information $I_{\rho}(A)$ when $\alpha=\frac{1}{2}$.

Denote $\mathcal{H}$ the $d$-dimensional complex Hilbert space. Let $\mathcal{C(H)}$ and $\mathcal{D(H)}$ be the sets of all complex and density matrices in $\mathcal{H}$, respectively. We use $\langle X,Y\rangle=\mathrm{Tr}(X^{\dag}Y)$ to denote the Hilbert-Schmidt inner product of $X,Y\in\mathcal{C(H)}$. Fix an orthogonal basis
$\{|i\rangle,i=1,2,\cdots,d\}$ of a $d$-dimensional Hilbert space
$\mathcal{H}$. It is not difficult to verify that $\{E_{ij}=|i\rangle\langle j|,\,i,j=1,2,\cdots,d\}$ is an
orthogonal basis for all observables in $\mathcal{C(H)}$, satisfying the orthogonal relations
$\langle E_{ij},E_{kl}\rangle=\mathrm{Tr}(E_{ji}E_{kl})=\delta_{ik}\delta_{jl}$. Accordingly, for any matrix $M=[m_{ij}]\in\mathcal{C(H)}$, $\vec{M}=(m_{11},m_{12},\cdots,m_{1d},m_{21},\cdots,m_{dd})^T$ can be expanded as its $d^2$-dimensional coordinate vector under this standard orthogonal basis $\{E_{ij}\}$. For any observables $A$, one has
\begin{align}\label{eq13}
A=&\sum_{i,j=1}^da_{ij}E_{ij}=\langle \vec{A},\vec{E}\rangle=\vec{A}^{\dag}\cdot\vec{E},
\end{align}
where $a_{ij}=\langle E_{ij},A\rangle=\mathrm{Tr}(E_{ji}A)$ and $\vec{E}=(E_{11},E_{12},\cdots, E_{1d},E_{21},\cdots,E_{dd})^T$.

The metric adjusted skew information of observable $A$ can be also written as \cite{HJ},
\begin{align}\label{eq14}
I_{\rho}^{c}(A)
=&\frac{m(c)}{2}K_{\rho}^{c}(\mathrm{i}[\rho,A],\mathrm{i}[\rho,A])\notag \\
=&\frac{m(c)}{2}\sum_{i,j=1}^d\sum_{k,l=1}^da_{ij}^{\ast}a_{kl}K_{\rho}^{c}
(\mathrm{i}[\rho,E_{ij}],\mathrm{i}[\rho,E_{kl}])\notag \\
=&\vec{A}^{\dag}\cdot\Gamma\cdot \vec{A},
\end{align}
where $\Gamma=(\Gamma_{i_j,k_l})$ and $\Gamma_{i_j,k_l}=\frac{m(c)}{2}K_{\rho}^{c}(\mathrm{i}[\rho,E_{ij}],
\mathrm{i}[\rho,E_{kl}])=Corr_{\rho}^c(E_{ij},E_{kl})$ with $i_j=d(i-1)+j$ and $k_l=d(k-1)+l$. In particular, when $c(x,y)=c_\alpha(x,y)$, $\Gamma_{i_j,k_l}=-\frac{1}{2}\mathrm{Tr}[\rho^\alpha,E_{ij}^{\dag}]
\left[\rho^{1-\alpha},E_{kl}\right]$\cite{HJ}. Since $\Gamma$ is semi-positive, there exists a matrix $O$ such that $\Gamma=O^{\dag}O$. Therefore, $I_{\rho}^{c}(A)=\vec{A}^{\dag}O^{\dag}O\vec{A}=(O\vec{A})^{\dag}(O\vec{A})=\|\vec{\alpha}\|^2$, where $\vec{\alpha}=O\vec{A}=(\alpha_1,\alpha_2,\cdots,\alpha_{d^2})^{\dag}$. Similarly, for observable $B$, we have $I_{\rho}^{c}(B)=\|\vec{\beta}\|^2$, with $\vec{\beta}=O\vec{B}=(\beta_1,\beta_2,\cdots,\beta_{d^2})^{\dag}$.
Therefore, for two arbitrary observables $A$ and $B$ the product form uncertainty relations via metric-adjusted skew information have the following form,
\begin{align}\label{eq15}
I_{\rho}^{c}(A)I_{\rho}^{c}(B)
=&\|\vec{\alpha}\|^2\|\vec{\beta}\|^2=\sum_{i,j=1}^{d^2}|\alpha_i|^2|\beta_j|^2    \notag \\
\geq& \left|\sum_{i,j=1}^{d^2}{\alpha^\ast_i} \beta_j \right|^2=|( \vec{\alpha},\vec{\beta})|^2=|Corr_{\rho}^{c}(A,B)|^2.
\end{align}

In \cite{HLTG}, Li et al. presented the following sum uncertainty relations via metric-adjusted skew information for finite $N$ observables $A_1, A_2,\cdots,A_N$ ($N\geq2$)
\begin{align}\label{eq16}
\sum_{i=1}^{N}I_{\rho}^{c}(A_i)\geq LB=\mathop{\mathrm{max}}\{LB_1,LB_2,LB_3\},
\end{align}
where
\begin{align}\label{eq17}
LB_1=
&\frac{1}{MN+(N-2)L}\left\{\frac{2L}{N(N-1)}\left(\sum_{1\leq i<j\leq N}\sqrt{I_{\rho}^{c}(A_i+A_j)}\right)^2+\right.
\nonumber\\
&\left.M\sum_{1\leq i<j\leq N}I_{\rho}^{c}(A_i-A_j)+
(M-L)I_{\rho}^{c}
\left(\sum_{i=1}^{N}A_{i}\right)\right\},
\end{align}
\begin{align}\label{eq18}
LB_2=
&\frac{1}{MN+(N-2)L}\left\{\frac{2M}{N(N-1)}\left(\sum_{1\leq i<j\leq N}\sqrt{I_{\rho}^{c}(A_i-A_j)}\right)^2+\right.
\nonumber\\
&\left.L\sum_{1\leq i<j\leq N}I_{\rho}^{c}(A_i+A_j)+
(M-L)I_{\rho}^{c}
\left(\sum_{i=1}^{N}A_{i}\right)\right\},
\end{align}
\begin{align}\label{eq19}
LB_3=
&\frac{1}{MN+(N-2)L}\left\{\frac{M-L}{(N-1)^2}\left(\sum_{1\leq i<j\leq N}\sqrt{I_{\rho}^{c}(A_i+A_j)}\right)^2 +\right.
\nonumber\\
&\left.
M\sum_{1\leq i<j\leq N}I_{\rho}^{c}(A_i-A_j)+L\sum_{1\leq i<j\leq N}I_{\rho}^{c}(A_i+A_j)
\right\},
\end{align}
the parameters $L$ and $M$ in (\ref{eq17}), (\ref{eq18}) and (\ref{eq19}) satisfy $M\geq L>0$, $L\geq M>0$ and $L>M>0$, respectively.

  For arbitrary observable $A_i$, ($i=1,\cdots,N$), $I_{\rho}^c(A_i)$ can be written as: $I_{\rho}^c(A_i)=\|\vec X_i\|^2$, where $\vec X_i=(x_{i1},x_{i2},\cdots,x_{id^2})^T$ with $x_{ij}=|\alpha_{ij}|$. Therefore, we have the following summation form uncertainty relations via metric-adjusted skew information.

{\bf Theorem 1} For finite $N$ observables $A_1, A_2, \cdots, A_N$ ($N\geq2$), the following sum uncertainty relations via metric-adjusted skew information hold,
\begin{align}\label{eq20}
\sum_{i=1}^{N}I_{\rho}^{c}(A_i)\geq \mathop{\mathrm{max}}\{LB1,LB2,LB3\},
\end{align}
where
\begin{align}\label{eq21}
LB1=
&\mathop{\mathrm{max}}\limits_{\pi_{i},\pi_{j}\in S_{d^2}}\frac{1}{MN+(N-2)L}\left\{\frac{2L}{N(N-1)}\left(\sum_{1\leq i<j\leq N}\Lambda_{\pi_{i}(i)\pi_{j}(j)}\right)^2+\right.
\nonumber\\
&\left.M\sum_{1\leq i<j\leq N}\overline{\Lambda}_{\pi_{i}(i)\pi_{j}(j)}^2+
(M-L)\Lambda_{\pi_{1}(1)\pi_{2}(2)\cdots\pi_{N}(N)}^2\right\},
\end{align}
\begin{align}\label{eq22}
LB2=
&\mathop{\mathrm{max}}\limits_{\pi_{i},\pi_{j}\in S_{d^2}}\frac{1}{MN+(N-2)L}\left\{\frac{2M}{N(N-1)}\left(\sum_{1\leq i<j\leq N}\overline{\Lambda}_{\pi_{i}(i)\pi_{j}(j)}\right)^2+\right.
\nonumber\\
&\left.L\sum_{1\leq i<j\leq N}\Lambda_{\pi_{i}(i)\pi_{j}(j)}^2+
(M-L)\Lambda_{\pi_{1}(1)\pi_{2}(2)\cdots\pi_{N}(N)}^2\right\},
\end{align}
\begin{align}\label{eq23}
LB3=
&\mathop{\mathrm{max}}\limits_{\pi_{i},\pi_{j}\in S_{d^2}}\frac{1}{MN+(N-2)L}\left\{\frac{M-L}{(N-1)^2}\left(\sum_{1\leq i<j\leq N}\Lambda_{\pi_{i}(i)\pi_{j}(j)}\right)^2 +\right.
\nonumber\\
&\left.
M\sum_{1\leq i<j\leq N}\overline{\Lambda}_{\pi_{i}(i)\pi_{j}(j)}^2+L\sum_{1\leq i<j\leq N}\Lambda_{\pi_{i}(i)\pi_{j}(j)}^2
\right\},
\end{align}
and
\begin{align*}
\Lambda_{\pi_{i}(i)\pi_{j}(j)}^2=&\sum_{k=1}^{d^2}(x_{i,\pi_{i}(k)}+x_{j,\pi_{j}(k)})^2,\\
\overline{\Lambda}_{\pi_{i}(i)\pi_{j}(j)}^2=&\sum_{k=1}^{d^2}(x_{i,\pi_{i}(k)}-x_{j,\pi_{j}(k)})^2,\\
\Lambda_{\pi_{1}(1)\pi_{2}(2)\cdots\pi_{N}(N)}^2=&\sum_{k=1}^{d^2}(x_{1,\pi_{1}(k)}+x_{2,\pi_{2}(k)}\cdots +x_{N,\pi_{N}(k)})^2,
\end{align*}
with $\pi_{i},\pi_{j}\in S_{d^2}$ the arbitrary $d^2$-element permutations, and $L$, $M$ in (\ref{eq21}), (\ref{eq22}) and (\ref{eq23}) satisfying $M\geq L>0$, $L\geq M>0$ and $L>M>0$, respectively.

\noindent\textit{Proof}
By using equations
\begin{align*}
\sum_{1\leq i<j\leq N} \|\vec X_i^{\pi_i}+\vec X_j^{\pi_j}\|^2=\left\| \sum_{i=1}^{N}\vec X_i^{\pi_i}\right\|^2+(N-2)\sum_{i=1}^{N}\|\vec X_i^{\pi_i}\|^2
\end{align*}
and
\begin{align*}
\sum_{1\leq i<j\leq N} \|\vec X_i^{\pi_i}-\vec X_j^{\pi_j}\|^2=N\sum_{i=1}^{N}\|\vec X_i^{\pi_i}\|^2-\left\| \sum_{i=1}^{N}\vec X_i^{\pi_i}\right\|^2,
\end{align*}
we derive that
\begin{align*}
\sum_{i=1}^{N} \|\vec X_i\|^2
&=\frac{1}{MN+(N-2)L}\left\{(M-L)\left\| \sum_{i=1}^{N}\vec X_i^{\pi_i}\right\|^2+\right.
\nonumber\\
&\left.M\sum_{1\leq i<j\leq N}\|\vec X_i^{\pi_i}-\vec X_j^{\pi_j}\|^2+L\sum_{1\leq i<j\leq N}\| \vec X_i^{\pi_i}+\vec X_j^{\pi_j}\|^2
\right\}
\end{align*}
for arbitrary $M, L\neq0$ holds.

By using Cauchy-Schwarz inequality,
\begin{align*}
\sum_{1\leq i <j\leq N}\|\vec X_i^{\pi_i}\pm \vec X_j^{\pi_j}\|^{2}\geq\frac{2}{N(N-1)}\left(\sum_{1\leq i <j\leq N}\|\vec X_i^{\pi_i}\pm \vec X_j^{\pi_j}\|\right)^2,
\end{align*}
we have
\begin{align*}
\sum_{i=1}^{N} \|\vec X_i^{\pi_i}\|^2
&\geq\frac{1}{MN+(N-2)L}\left\{\frac{2L}{N(N-1)}\left(\sum_{1\leq i<j\leq N}\| \vec X_i^{\pi_i}+\vec X_j^{\pi_j}\|\right)^2+\right.
\nonumber\\
&\left.M\sum_{1\leq i<j\leq N}\|\vec X_i^{\pi_i}-\vec X_j^{\pi_j}\|^2+
(M-L)\left\|\sum_{i=1}^{N}\vec X_i^{\pi_i}\right\|^2\right\}
\end{align*}
and
\begin{align*}
\sum_{i=1}^{N} \|\vec X_i^{\pi_i}\|^2
&\geq\frac{1}{MN+(N-2)L}\left\{\frac{2M}{N(N-1)}\left(\sum_{1\leq i<j\leq N}\| \vec X_i^{\pi_i}-\vec X_j^{\pi_j}\|\right)^2+\right.
\nonumber\\
&\left.L\sum_{1\leq i<j\leq N}\|\vec X_i^{\pi_i}+\vec X_j^{\pi_j}\|^2+
(M-L)\left\|\sum_{i=1}^{N}\vec X_i^{\pi_i}\right\|^2\right\}
\end{align*}
for $M, L\geq0$. When $L>M>0$, due to $\|\sum_{i=1}^{N}\vec X_i^{\pi_i}\|^2\leq\frac{1}{(N-1)^2}(\sum_{1\leq i<j\leq N}\| \vec X_i^{\pi_i}+\vec X_j^{\pi_j}\|)^2$, we obtain
\begin{align*}
\sum_{i=1}^{N} \|\vec X_i^{\pi_i}\|^2
&\geq\frac{1}{MN+(N-2)L}\left\{\frac{M-L}{(N-1)^2}\left(\sum_{1\leq i<j\leq N}\|\vec X_i^{\pi_i}+\vec X_j^{\pi_j}\|\right)^2+\right.
\nonumber\\
&\left.M\sum_{1\leq i<j\leq N}\|\vec X_i^{\pi_i}-\vec X_j^{\pi_j}\|^2+L\sum_{1\leq i<j\leq N}\| \vec X_i^{\pi_i}+\vec X_j^{\pi_j}\|^2
\right\},
\end{align*}
where $\vec X_i^{\pi_i}=(x_{i,{\pi_i}(1)},x_{i,{\pi_i}(2)},\cdots,x_{i,{\pi_i}(d^2)})^T$.
This completes the proof. $\Box$

When $\pi_{i},\pi_{j}\in S_{d^2}$ in Theorem 1 are just the identity permutations, we have the following Corollary.

{\bf Corollary 1} For finite $N$ observables $A_1, A_2, \cdots, A_N$ ($N\geq2$), the following sum uncertainty relations via metric-adjusted skew information hold,
\begin{align}\label{eq24}
  \sum_{i=1}^{N}I_{\rho}^{c}(A_i)\geq\overline{LB}=\mathop{\mathrm{max}}\{\overline{LB_1},\overline{LB_2},\overline{LB_3}\}.
\end{align}
where
\begin{align}\label{eq25}
\overline{LB_1}=
&\frac{1}{MN+(N-2)L}\left\{\frac{2L}{N(N-1)}\left[\sum_{1\leq i<j\leq N}
 \sqrt {\sum_{k=1}^{d^2}(x_{ik}+x_{jk})^2}\right]^2+\right.
\nonumber\\
&\left.M\sum_{1\leq i<j\leq N}\sum_{k=1}^{d^2}(x_{ik}-x_{jk})^2+
(M-L)\sum_{k=1}^{d^2}(x_{1k}+x_{2k}+\cdots x_{Nk})^2\right\},
\end{align}
\begin{align}\label{eq26}
\overline{LB_2}=
&\frac{1}{MN+(N-2)L}\left\{\frac{2M}{N(N-1)}\left[\sum_{1\leq i<j\leq N}
 \sqrt {\sum_{k=1}^{d^2}(x_{ik}-x_{jk})^2}\right]^2+\right.
\nonumber\\
&\left.L\sum_{1\leq i<j\leq N} \sum_{k=1}^{d^2}(x_{ik}+x_{jk})^2+
(M-L)\sum_{k=1}^{d^2}(x_{1k}+x_{2k}+\cdots x_{Nk})^2 \right\},
\end{align}
\begin{align}\label{eq27}
\overline{LB_3}=
&\frac{1}{MN+(N-2)L}\left\{\frac{M-L}{(N-1)^2}\left[\sum_{1\leq i<j\leq N}
 \sqrt {\sum_{k=1}^{d^2}(x_{ik}+x_{jk})^2}\right]^2+\right.
\nonumber\\
&\left.
M\sum_{1\leq i<j\leq N} \sum_{k=1}^{d^2}(x_{ik}-x_{jk})^2+L\sum_{1\leq i<j\leq N} \sum_{k=1}^{d^2}(x_{ik}+x_{jk})^2
\right\},
\end{align}
with the parameters $L$, $M$ in (\ref{eq25}), (\ref{eq26}) and (\ref{eq27}) satisfying $M\geq L>0$, $L\geq M>0$ and $L>M>0$, respectively.

Next we present two examples via WYD skew information to illustrate that our result is tighter than the existing one. As a special case, we take $M=2$, $L=1$ for inequalities (\ref{eq17}) and (\ref{eq25}), $M=1$, $L=2$ in inequalities (\ref{eq18}), (\ref{eq19}), (\ref{eq26}) and (\ref{eq27}).

{\bf Example 1} Let $\rho=\frac{1}{2}(\mathrm{I_2}+\mathbf{r}\cdot\bm{\sigma})$ be a mixed state with $\mathbf{r}=(\frac{3}{4}\sin\theta,0,\frac{3}{4}\cos\theta)$, where $\bm{\sigma}=(\sigma_x,\sigma_y,\sigma_z)$ is the vector of Pauli matrices, $\mathrm{I_2}$ is the $2\times2$ identity operator. Consider observables
\begin{equation*}
A_1=
\begin{pmatrix}
1& 2+\mathrm{i}\\
2-\mathrm{i}& -1
\end{pmatrix},\quad
A_2=
\begin{pmatrix}
1& \mathrm{i}\\
-\mathrm{i} & -1
\end{pmatrix},
\end{equation*}
\begin{equation*}
A_3=
\begin{pmatrix}
0& 1+\frac{\mathrm{i}}{2}\\
1-\frac{\mathrm{i}}{2}& 0
\end{pmatrix},\quad
A_4=
\begin{pmatrix}
0& \mathrm{i}\\
-\mathrm{i} & 0
\end{pmatrix}.
\end{equation*}
\begin{figure}[ht]\centering
\subfigure[]
{\begin{minipage}[Xu-Cong-uncertaintyMaSI2a]{0.49\linewidth}
\includegraphics[width=0.95\textwidth]{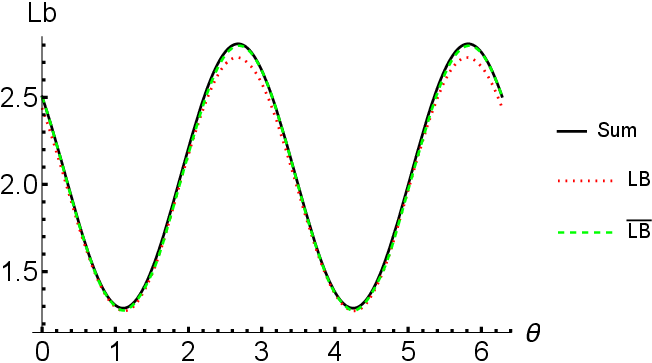}
\end{minipage}}
\subfigure[]
{\begin{minipage}[Xu-Cong-uncertaintyMaSI2b]{0.49\linewidth}
\includegraphics[width=0.95\textwidth]{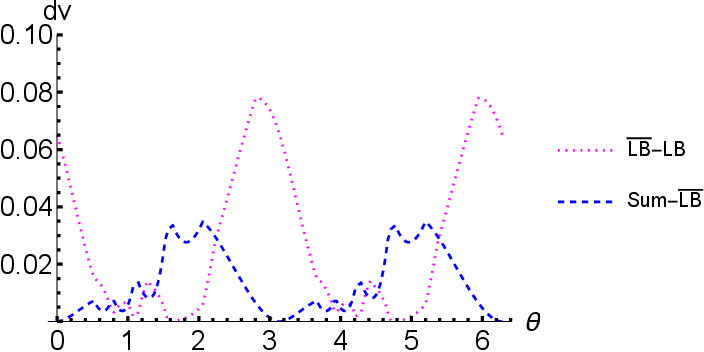}
\end{minipage}}
\caption{{The solid black, dotted red, dashed green curves represent the sum$=I_{\rho}^{\frac{1}{3}}(A_1)+I_{\rho}^{\frac{1}{3}}(A_2)+I_{\rho}^{\frac{1}{3}}(A_3)+I_{\rho}^{\frac{1}{3}}(A_4)$, the lower bounds (Lb) $LB$ and $\overline{LB}$, respectively. (b) The dotted magenta, dashed blue curves denotes the difference value (dv) between $\overline{LB}$ and $LB$, sum and $\overline{LB}$. \label{fig:Fig1}}}
\end{figure}
Under the standard orthogonal basis $\{e_1=E_{11}, e_2=E_{12}, e_3=E_{21}, e_4=E_{22}\}$, we have
$\Gamma_{i,j}=-\frac{1}{2}\mathrm{Tr}([\rho^{\frac{1}{3}},e_i^{\dag}][\rho^{\frac{2}{3}},e_{j}])$, where $c(x,y)=c_{\alpha}(x,y)$ with $\alpha=\frac{1}{3}$.
We obtain $\vec A_i$ and $x_{ij}$ accordingly. The lower bound $LB$ and $\overline{LB}$ can be calculated from (\ref{eq16}) and (\ref{eq24}). They are shown in Figure $1$ $(a)$. Their differences are shown in Figure $1$ $(b)$.
Figure $1$ shows that our lower bound is tighter than the lower bound $LB$ given in \cite{HLTG}.

{\bf Example 2} Consider state $\rho=\frac{1}{2}(\mathrm{I_2}+\mathbf{r}\cdot\bm{\sigma})$ with $\mathbf{r}=(\frac{\sqrt{2}}{3},0,\frac{\sqrt{2}}{3}\cos\theta)$, and observables $A'_1=\sigma_x$, $A'_2=\sigma_y$ and $A'_3=\sigma_z$.
\begin{figure}[ht]\centering
\subfigure[]
{\begin{minipage}[Xu-Cong-uncertaintyMaSI3a]{0.49\linewidth}
\includegraphics[width=0.95\textwidth]{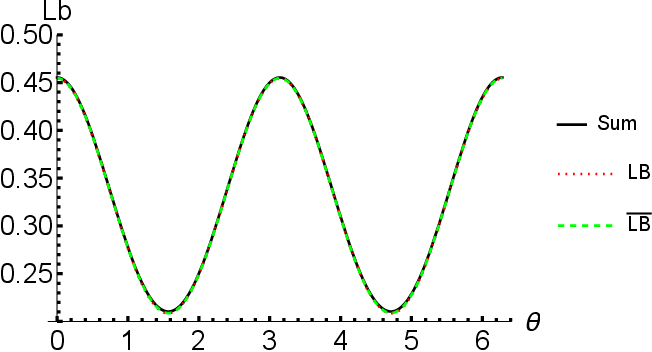}
\end{minipage}}
\subfigure[]
{\begin{minipage}[Xu-Cong-uncertaintyMaSI3b]{0.49\linewidth}
\includegraphics[width=0.95\textwidth]{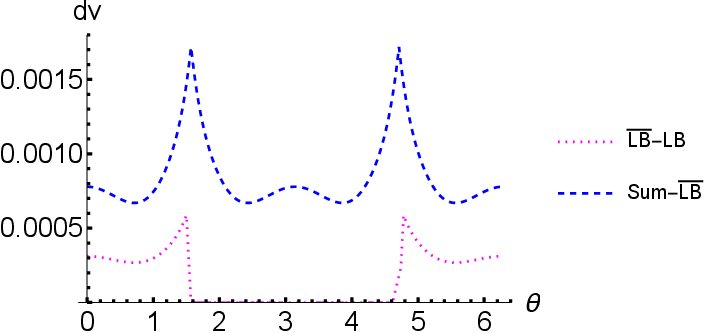}
\end{minipage}}
\caption{{The solid black, dotted red, dashed green curves represent the sum$=I_{\rho}^{\frac{1}{3}}(A'_1)+I_{\rho}^{\frac{1}{3}}(A'_2)+I_{\rho}^{\frac{1}{3}}(A'_3)$, the lower bounds (Lb) $LB$ and $\overline{LB}$, respectively. (b)The dotted magenta, dashed blue curves denote the difference value (dv) between $\overline{LB}$ and $LB$, sum and $\overline{LB}$. \label{fig:Fig2}}}
\end{figure}
Under the standard orthogonal basis $\{e_1=\frac{1}{\sqrt{2}}\mathrm{I_2},e_2=\frac{1}{\sqrt{2}}\sigma_x, e_3=\frac{1}{\sqrt{2}}\sigma_y, e_4=\frac{1}{\sqrt{2}}\sigma_z\}$, we correspondingly have $\Gamma_{i,j}=-\frac{1}{2}\mathrm{Tr}([\rho^{\frac{1}{3}},e_i^{\dag}][\rho^{\frac{2}{3}},e_{j}])$, where $c(x,y)=c_{\alpha}(x,y)$ with $\alpha=\frac{1}{3}$. The lower bound $LB$ and $\overline{LB}$ are calculated from (\ref{eq16}) and (\ref{eq24}),
see Figure $2$ $(a)$, and Figure $2$ $(b)$ for their differences. Figure $2$ shows that our bound is tighter than the lower bound $LB$ again.
FIG. 1 (2) implies that under the standard orthogonal basis $\{e_1=E_{11}, e_2=E_{12}, e_3=E_{21}, e_4=E_{22}\}$ ($\{e_1=\frac{1}{\sqrt{2}}\mathrm{I_2},e_2=\frac{1}{\sqrt{2}}\sigma_x,e_3=\frac{1}{\sqrt{2}}\sigma_y, e_4=\frac{1}{\sqrt{2}}\sigma_z\}$) the lower bound $LB$ and $\overline{LB}$ are all close to the sum $=I_{\rho}^{\frac{1}{3}}(A_1)+I_{\rho}^{\frac{1}{3}}(A_2)+I_{\rho}^{\frac{1}{3}}(A_3)
+I_{\rho}^{\frac{1}{3}}(A_4)$($I_{\rho}^{\frac{1}{3}}(A'_1)
+I_{\rho}^{\frac{1}{3}}(A'_2)+I_{\rho}^{\frac{1}{3}}(A'_3)$), the lower bound $\overline{LB}$ is tighter than the one in \cite{HLTG}.

\noindent {\bf 3. Product form uncertainty relations via metric-adjusted skew information}

Let $\vec X=(x_1,x_2,\cdots,x_{d^2})^T$ ($\vec Y=(y_1,y_2,\cdots,y_{d^2})^T$), and $x_i=|\alpha_i|$ ($y_i=|\beta_i|$) be non negative real number.
Ma et al. defined the quantity $I_k(A,B)$ based on metric-adjusted skew information \cite{MZF} as,
\begin{align}\label{eq28}
I_k(A,B)=&\sum_{1\leq i\leq d^2}x_i^2y_i^2+\sum_{\stackrel{1\leq i<j\leq d^2}{k<j}}(x_i^2y_j^2+x_j^2y_i^2)+\sum_{1\leq i<j\leq k}2x_iy_ix_jy_j.
\end{align}
Note that $1\leq k\leq d^2$, $I_1(A,B)=I_{\rho}^c(A)I_{\rho}^c(B)=\|\vec{X}\|^2\|\vec{Y}\|^2$ and
\begin{align}\label{eq29}
I_{k+1}(A,B)-I_k(A,B)=-\left(\sum_{i=1}^kx_iy_{k+1}-y_ix_{k+1}\right)^2\leq0,
\end{align}
namely, $I_1(A,B)\geq I_2(A,B)\geq\cdots \geq I_{d^2}(A,B)$,
where $I_2(A,B)$ is the optimal lower bound. When $k=2$, we have
\begin{align}\label{eq30}
I_2(A,B)=\sum_{i,j=1}^{d^2}x_i^2y_j^2-x_1^2y_2^2-x_2^2y_1^2+2x_1x_2y_1y_2.
\end{align}

Inspired by the inequality $\Delta A^2\Delta B^2\geq I'_1(A,B)$ in Theorem $1$ \cite{JSZ}, where $I'_1(A,B)$ is the lower bound via variance of operator $A$ and $B$. We have the following product form uncertainty relations based on metric-adjusted skew information.

{\bf Theorem 2} For two arbitrary observables $A$, $B$ on a $d$-dimensional ($d\geq3$) Hilbert space $\mathcal{H}$. The product of metric-adjusted skew information of $A$ and $B$ satisfies the following uncertainty relations,
\begin{align}\label{eq31}
I_{\rho}^c(A)I_{\rho}^c(B)\geq \mathop{\mathrm{max}}\limits_{\pi_1,\pi_2\in S_{d^2}}(\pi_1,\pi_2)\widetilde{I}'_1(A,B),
\end{align}
where
\begin{align}\label{eq32}
(\pi_1,\pi_2)\widetilde{I}'_1(A,B)=\sum_{i=1}^{d^2}x^2_{{\pi_1}(i)} y^2_{{\pi_2}(i)}+\sum_{\stackrel{i\neq j}{i\neq1}}^{d^2} x^2_{{\pi_1}(i)}y^2_{{\pi_2}(j)}+x^2_{{\pi_1}(1)}\sum_{j=4}^{d^2}y^2_{{\pi_2}(j)}+2x^2_{{\pi_1}(1)}y_{{\pi_2}(2)}y_{{\pi_2}(3)},
\end{align}
$\pi_1$ and $\pi_2$ stand for arbitrary $d^2$ element permutations in symmetry group $S_{d^2}$ and the equality holds if and only if $x_{{\pi_1}(1)}y_{{\pi_2}(2)}=x_{{\pi_1}(1)}y_{{\pi_2}(3)}$.

{\it Proof}
\begin{align}\label{eq33}
I_{\rho}^{c}(A)I_{\rho}^{c}(B)
=&\sum_{i,j=1}^{d^2}x^2_{{\pi_1}(i)}y^2_{{\pi_2}(j)} \\ \notag
=&\sum_{i=1}^{d^2}x^2_{{\pi_1}(i)}y^2_{{\pi_2}(i)}+\sum_{\stackrel{i\neq j}{ i\neq1}}^{d^2}x^2_{{\pi_1}(i)}y^2_{{\pi_2}(j)}+x^2_{{\pi_1}(1)}\sum_{j=2}^{d^2}y^2_{{\pi_2}(j)}\\ \notag
\geq&\sum_{i=1}^{d^2}x^2_{{\pi_1}(i)}y^2_{{\pi_2}(i)}+\sum_{\stackrel{i\neq j}{i\neq1}}^{d^2}x^2_{{\pi_1}(i)}y^2_{{\pi_2}(j)}+x^2_{{\pi_1}(1)}\sum_{j=4}^{d^2}y^2_{{\pi_2}(j)}+2x^2_{{\pi_1}(1)}y_{{\pi_2}(2)}y_{{\pi_2}(3)}\\ \notag
=&(\pi_1,\pi_2)\widetilde{I}'_1(A,B).
\end{align}

We have the following Corollary when $\pi_1$ and $\pi_2$ in Theorem 2 are identity permutations.

{\bf Corollary 2} For two arbitrary observables $A$, $B$, we have the following uncertainty relations:
\begin{align}\label{eq34}
I_{\rho}^c(A)I_{\rho}^c(B)\geq \widetilde{I}'_1(A,B),
\end{align}
where
\begin{align}\label{eq35}
\widetilde{I}'_1(A,B)=\sum_{i=1}^{d^2}x_i^2y_i^2+\sum_{\stackrel{i\neq j}{ i\neq1}}^{d^2} x_i^2y_j^2+x_1^2\sum_{j=4}^{d^2}y_j^2+2x_1^2y_2y_3.
\end{align}

The lower bound in Corollary 2 is tighter than the lower bound $I_{\rho}^c(A)I_{\rho}^c(B)\geq I_k(A,B)$ in Theorem 1 \cite{MZF} for particular cases. In fact, the right hands of the above theorems and corollaries via metric-adjusted skew information depend not only on the quantum $\rho$ but also on the choice of the basis. Because the observable has different coordinate vector with respect to the different standard basis. The metric-adjusted skew information can not be calculated in general. We take $c(x,y)=c_{\alpha}(x,y)$ or other Morozova-Chentsov function. In the following example, we compare the lower bound in Corollary 2 and $I_2(A,B)$ for WYD skew information.

{\bf Example 3} Let $\rho=\frac{1}{2}(\mathrm{I_2}+\mathbf{r}\cdot\bm{\sigma})$ be a mixed (pure) state with $\mathbf{r}=(\frac{2}{3}\sin\theta,0,\frac{2}{3}\cos\theta)$, ($\mathbf{r}=(\frac{2}{3}\sin\theta,\frac{\sqrt{5}}{3},\frac{2}{3}\cos\theta)$). Consider observables $A=\sigma_x$ and $B=\sigma_y$.
\begin{figure}[ht]\centering
\subfigure[]
{\begin{minipage}[Xu-Cong-uncertaintyMaSI1a]{0.49\linewidth}
\includegraphics[width=0.95\textwidth]{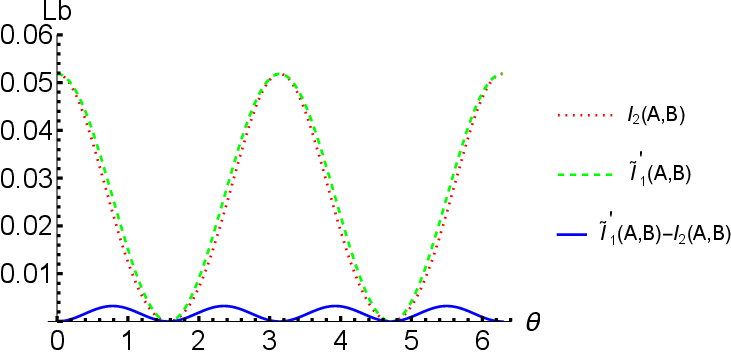}
\end{minipage}}
\subfigure[]
{\begin{minipage}[Xu-Cong-uncertaintyMaSI1b]{0.49\linewidth}
\includegraphics[width=0.95\textwidth]{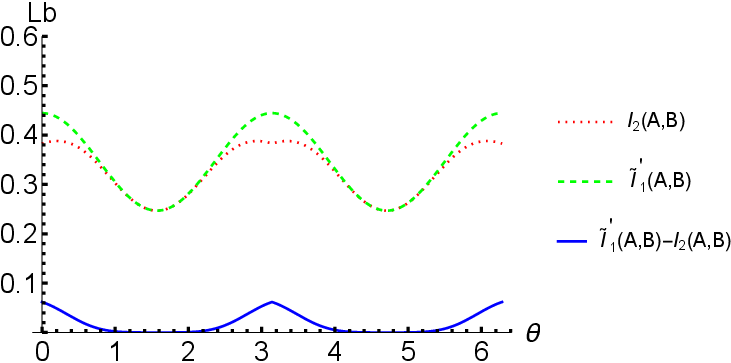}
\end{minipage}}
\caption{{The dotted red, dashed green and solid blue curves represent the lower bounds (Lb) $I_2(A,B)$, $\widetilde{I}'_1(A,B)$ and the difference $\widetilde{I}'_1(A,B)-I_2(A,B)$, respectively. (a) ((b)) for mixed (pure) state: under the standard orthogonal basis
$\{e_1=E_{11}, e_2=E_{12}, e_3=E_{21}, e_4=E_{22}\}$, we have $\Gamma_{i,j}=-\frac{1}{2}\mathrm{Tr}([\rho^{\frac{1}{3}},e_i^{\dag}][\rho^{\frac{2}{3}},e_{j}])$, where $c(x,y)=c_{\alpha}(x,y)$ with $\alpha=\frac{1}{3}$.\label{fig:Fig3}}}
\end{figure}
From FIG. 3, we can see that the lower bound in Corollary 2 is tighter than $I_2(A,B)$ under the standard orthogonal basis
$\{e_1=E_{11}, e_2=E_{12}, e_3=E_{21}, e_4=E_{22}\}$. In fact, if we choose the standard orthogonal basis
$\{e_1=\frac{1}{\sqrt{2}}\mathrm{I_2},e_2=\frac{1}{\sqrt{2}}\sigma_x, e_3=\frac{1}{\sqrt{2}}\sigma_y, e_4=\frac{1}{\sqrt{2}}\sigma_z\}$
for the mixed (pure) state in Example $3$ , through calculation we find that $x_1=y_1=0$ in (\ref{eq30}) and (\ref{eq35}), i.e., $I_2(A,B)=\widetilde{I}'_1(A,B)=I_{\rho}^{\frac{1}{3}}(A)I_{\rho}^{\frac{1}{3}}(B)$.
The Example 3 implies that the lower bounds of Corollary $2$ is tighter than the result in Theorem 1 \cite{MZF}.

\vskip0.1in

\noindent {\bf 4. Conclusions}\\\hspace*{\fill}\\
By using sampling observables coordinates, the tighter sum form uncertainty relations of observables have obtained. By explicit examples, we have shown that our lower bound is tighter than the results in \cite{HLTG}. we have also presented the product form uncertainty relations via metric-adjusted skew information for arbitrary two observables. These lower bounds depend on the choose of the standard orthogonal basis. By concrete examples, we have shown that our results are better than the previous ones \cite{MZF}. The results and approach we used may be helpful for future researches on uncertainty relations via more generalized skew information.

\vskip0.1in

\noindent
\subsubsection*{Acknowledgements}
This work was supported by National Natural Science Foundation of
China (Grant Nos. 12161056, 12075159, 12171044); the Academician
Innovation Platform of Hainan Province; and Changsha University of Science
and Technology (Grant No. 097000303923).


\subsubsection*{Competing interests}
\small {The authors declare no competing interests.}



\end{document}